\documentclass[11pt,a4 paper]{article}

\usepackage{amssymb}
\usepackage{amsmath}
\usepackage[dvips]{graphicx}
\usepackage{subfig}
\usepackage{multirow}
\usepackage{array}

\begin{document}
\thispagestyle{empty}
\begin{center}
\noindent {\textbf{\large Friedmann cosmology with particle creation in modified $f(R,T)$ gravity}}

\vspace{1cm}

\noindent \textbf {C. P. Singh\footnote{Corresponding author}} and \textbf{Vijay Singh$^2$}

\vspace{0.5cm}

\noindent $^{1,2}$Department of Applied Mathematics,\\
 Delhi Technological University\\
 (Formerly Delhi College of
 Engineering)\\
 Bawana Road, Delhi-110 042, India.\\
{ $^1$cpsphd@rediffmail.com\\
$^2$gtrcosmo@gmail.com}
\end{center}

\vspace{1.5cm}

\noindent{\textbf{Abstract}}\\
The theoretical and observational consequences of thermodynamics of open systems, which allow particle creation are investigated in modified $f(R,T)$ ($R$ is the Ricci scalar and $T$ is the trace of energy-momentum tensor) theory of gravity within the framework of a flat Friedmann-Robertson-Walker line element. A simplest particular model $f(R,T)=R+2f(T)$ and ``gamma-law" equation of state $p=(\gamma-1)\rho$ are assumed to explore the role of particle creation in the early and late time evolution of the universe. A power-law expansion model is proposed for $0\leq \gamma \leq 2$ by considering the natural phenomenological particle creation rate $\psi=3\beta nH$, where $\beta$ is a pure number of the order of unity. A Big Rip singularity is observed for $\gamma<0$, which describes the phantom cosmology. We observe that the accelerated expansion of the universe is driven by the particle creation without any exotic ``quintessence" component or a cosmological constant. It is also found by density parameter that the model becomes of negative curvature due to particle creation and the entropy increases with the evolution of the universe. Some kinematics tests such as lookback time, luminosity distance, proper distance, angular diameter versus redshift are discussed in detail to observe the role of particle creation in early and late time evolution of the universe.\\

\noindent {\textbf{ Keywords:}} Modified gravity theory; Particle creation.\\

\noindent PACS number(s): 98.80-k, 98.80-cq, 04.20-q.\\
\pagebreak

\pagestyle{myheadings}
\noindent \textbf{1. Introduction}\\

\noindent Many theoretical and observational studies of the universe, such as Type Ia supernovae$^{1-3}$, cosmic microwave background (CMB) anisotropy$^4$, large scale structures (LSS)$^5$ etc., have been shown that a pure Friedmann-Lamaitre-Robertson-Walker (FLRW) cosmology with matter and radiation could not explain all the large scale properties of our universe. The observations suggest that our universe is spatially flat and consists of about 70 percent dark energy with negative pressure, nearly 25 percent of dark matter, which populates the galaxy halos and 5 percent of baryonic matter. The late time cosmic acceleration is usually assumed to be driven by a cosmic fluid/field generically known as dark energy (DE)$^{6-8}$, which can be observed only by its gravitational effects$^{9-11}$. Since the time of this discovery in 1998, a large number of probable candidates of DE have been proposed to explain late time acceleration of the universe. The two most accepted DE models are that of a cosmological constant$^{12}$  (assumed possibly to be the quantum vacuum energy) and a slowly rolling scalar field (quintessence models)$^{13-17}$. All of these DE cosmological models are based on the Big Bang cosmology. In the context of early universe, the standard Big Bang model presents theoretical and observational difficulties, such as the singularity problem$^{18}$, flatness and horizon problem, reheating during the inflationary epoch$^{19}$, confliction between the age of the universe and the age of the oldest stars in globular clusters (age problem)$^{20}$, the entropy problem$^{21,22}$ etc.\\
\indent The flatness and horizon problems together with the entropy problem have been resolved (up to a certain extent) by the inflationary model proposed by Guth$^{23}$ and Linde$^{24}$. The age confliction$^{25}$ is not an isolated complication, it comes with another serious one trouble that is structure formation through gravitational amplification of small primeval density perturbation. These issues open the door of investigations of different alternative theories of gravity for studying not only unpredicted late time cosmic acceleration but early universe as well $^{26-29}$. Many pioneer concepts have been introduced by theorists to alleviate the problems of early and late universes. Among the ways to resolve the problems of early universe, Dirac's large number hypothesis inspired a class of new cosmology named particle creation$^{30-32}$. The steady state model introduced by Bondi and Gold$^{33}$ on the foundation of perfect cosmological principle (PCP) also asserts the continuous generation of matter in the universe. Hoyle$^{34}$ and Narlikar$^{35}$ independently proposed a creation field theory and studied the matter creation during the evolution of the universe. Tryon$^{36}$ and Fomin$^{37}$ in their individual work proposed a theoretical concept of the creation of the universe as a vacuum fluctuation. Brout et al.$^{38-41}$ putted a strong foundation of simultaneous creation of matter and curvature from a quantum fluctuation of the Minkowskian space-time vacuum.\\
\indent Later on, Gunzig et al.$^{42}$ and Prigogine et al.$^{43}$ established the theoretical scenario of matter creation in the framework of cosmology. The authors applied the thermodynamics of an open system to cosmology allowing particle creation and entropy production. They showed that the second law of thermodynamics might be modified to accommodate the flow of energy from gravitational field to the matter field, resulting in the creation of particles and consequently entropy. The work might suggest that at the expense of the gravitational field, particle creation takes place as an irreversible process constrained by the usual requirements of the non-equilibrium thermodynamics, however, the reverse process (matter destruction) thermodynamically forbidden. Calv$\acute a$o et al.$^{44}$ extended this new concept of matter creation under adiabatic conditions. The further results were generalized by Lima and Germano$^{45}$ through a contravariant formulation allowing specific entropy variation as usually expected for non-equilibrium process in fluids. Lima and Alcaniz$^{46}$ and Alcaniz and Lima$^{47}$ investigated observational consequences of FRW models driven by adiabatic matter creation through some kinematics tests. After the discovery of accelerating universe the particle creation theory was reconsidered to explain it and got unexpected results. The negative pressure due to particle creation, might play the role of exotic matter component. Zimdahl et al.$^{48}$ tested the particle creation with SNe Ia data to get the result of accelerating universe. Qiang et al.$^{49}$ studied a model with adiabatic particle creation which is also consistent with SNe Ia data. Singh and Beesham$^{50,51}$, and  Singh$^{52,53}$ studied early universe in FRW cosmology with particle creation through some kinematical tests. This theoretical formulation of continuous creation of matter in the universe may reinterpret several predictions of the standard Big Bang cosmology.\\
\indent On the other hand, many modifications of general relativity (GR) have been proposed to reconcile the problems plagued in cosmology$^{54}$. Among such modifications the simplest one is replacing the Ricci scalar curvature $R$ with the general function of $R$ called $f(R)$ theories of gravity, describe transition from decelerated to accelerated expansion of the universe$^{55}$. Bertolami et al.$^{56}$ have generalized $f(R)$ theories by introducing an explicit coupling between arbitrary function of the Ricci scalar $R$ and the matter Lagrangian density. Harko$^{57}$ has extended this new concept to the arbitrary coupling between matter and geometry. Harko et al.$^{58}$ have considered another extension of GR, where the gravitational Lagrangian is given by an arbitrary function of the Ricci scalar $R$ and of the trace $T$ of the stress-energy tensor, so called $f(R,T)$ theory of gravity. The justification of choosing $T$ as an argument for the Lagrangian is from exotic imperfect fluids or quantum effects. The authors have argued that due to the coupling between matter and geometry the theory depends on a source term, representing the variation of the matter-stress-energy tensor with respect to the metric. Consequently, the cosmic acceleration in $f(R,T)$ gravity results not only from geometrical effect but also from the matter contribution. These interesting features of $f(R,T)$ gravity motivate many authors to study it for resolving several issues of current interest in cosmology and astrophysics.\\
\indent Jamil et al.$^{59}$ have reconstructed cosmological models in the framework of $f(R,T)$ gravity, showing dust fluid $\Lambda CDM$, phantom-non-phantom era and phantom cosmology, which are good agreement with BAO observational data for low-redshift $z<2$. Houndjo$^{60}$ has reconstructed $F(R,T)=f_1(R)+f_2(T)$ using auxiliary scalar field with two known examples of scale factors corresponding to an accelerating universe. Houndjo and Piatella$^{61}$ have numerically reconstructed $f(R,T)$ theory with dark matter and holographic DE, describing the same expansion history in GR. Houndjo et al.$^{62}$ have investigated $f(R,T)$ gravity models which reproduce four types of future finite-time singularities. Sharif and Zubair$^{63}$ have discussed a non-equilibrium thermodynamics in $f(R,T)$ gravity by taking two forms of the energy-momentum tensor of dark components which endorses second law of thermodynamics holds both in phantom and non-phantom phases. Jamil et al.$^{64}$ have proved, in general, the first law of black hole thermodynamics is violated in $f(R,T)$ gravity. Alvarenga et al.$^{65}$ have paid special attention on $f(R,T)=R+2f(T)$ assuming special function $f(T)$ showing energy conditions can be satisfied for suitable input parameters. Sharif et al.$^{66}$ and Sharif and Zubair$^{67}$ have derived energy conditions in the context of $f(R,T)$ gravity, which can reduce to the well-known conditions in $f(R)$ and GR. Chakraborty$^{68}$ has shown that a part of an arbitrary function $f(R,T)$ can be determined taking into account of the conservation of stress-energy tensor. Alvarenga et al.$^{69}$ have studied the evolution of scalar cosmological perturbations in the background of metric formalism in $f(R,T)$ theory. Pasqua et al.$^{70}$ have considered modified holographic Ricci DE (MHRDE) model in $f(R,T)$ modified gravity, which explained a quintessence-like behavior of the model. Shabani and Farhaudi$^{71}$ have studied the cosmological solutions of $f(R,T)$ gravity through phase space analysis. Sharif and Zubair$^{72}$ have considered $f(R,T)$ theory as an effective description for the holographic and new agegraphic DE and reconstructed the corresponding models, which explain phantom or quintessence regimes of the universe. Recently, Singh and Singh$^{73}$ have reconstructed the particular form $f(R,T)=R+2f(T)$ by assuming de-Sitter and power-law expansion of scale factors and showed the reconstructed forms of $f(R,T)$ could explain DE models successfully. However, all these mentioned works have been carried out in FRW spacetime. Some authors have also explored $f(R,T)$ theory in anisotropic background$^{74-77}$ and higher dimension$^{78}$. Thus, the newly proposed $f(R,T)$ gravity motivates us to study the particle creation phenomena in the early and late time evolution of the universe.\\
\indent  In this paper, we investigate the theoretical and observational significance of particle creation in modified $f(R,T)$ gravity theory in the framework of a flat Friedmann-Robertson-Walker model. Exact solutions of the field equations are obtained by assuming the suitable form of $f(R,T)=R+2f(T)$, equation of state, and particle creation rate. We obtain the non-singular power-law solution for the scale factor. A Big Rip singularity is also observed which describes the phantom cosmology. We study some kinematical tests to explain the physical significance of particle creation during early and late time evolution of the universe.\\
\indent The paper is organized as follows. The thermodynamics of particle creation is presented in the Sec. 2. A brief review of modified $f(R,T)$ gravity theory and its field equations are given in Sec. 3. In Sec. 4, the model and field equations are presented with particle creation.  The exact solutions of the field equations with suitable assumptions are obtained in Sec. 5. Sec. 6 and its subsections are devoted to study some kinematics tests of the model. Finally, the outcomes are summarized in Sec. 7.\\

\noindent\textbf{2. Thermodynamics of particle creation}\\

\noindent If we regard the whole universe as a closed thermodynamical system in which number of the particles in a given volume is constant, then the laws of thermodynamics  have the form
\begin{equation}
  d(\rho_m V)=dQ-p_mdV,
\end{equation}
\noindent and
\begin{equation}
  TdS=d(\rho_m V)+p_mdV,
\end{equation}
\noindent where $\rho_m$, $V$,  $p_m$, $T$ and $S$ are the energy density, volume, thermodynamical pressure, temperature, and entropy, respectively. Here, $dQ$ is the heat received by the system during time $dt$. It is observed from (1) and (2) that the entropy production is given by\\
\begin{equation}
 TdS=dQ.
\end{equation}
\noindent Eq. (3) shows that the entropy remains stationary, i.e.,  $dS=0$ for a closed adiabatic system for which $dQ=0$. However, if we treat the universe as an open thermodynamic system allowing irreversible matter creation from the energy of the gravitational field, we can account for entropy production right from the beginning, and the second law of thermodynamics is also incorporated into the evolutionary equations in a more meaningful way. Thus, the creation of matter acts as a source of internal energy. In such situation the number of particles $N$ in a given volume $V$ is not to be a constant but is time dependent. Therefore, Eq. (1) modifies as
\begin{equation}
  d(\rho_m V)=dQ-p_mdV+(h/n)\;d(nV),
\end{equation}
\noindent where $N=nV$, $n$ is the particle number density, and $h=(\rho_m+p_m)$ is the enthalpy per unit volume of the system. In case of adiabatic system where $dQ=0$, Eq. (4) for an open system reduces to
\begin{equation}
  d(\rho_m V)+p_mdV=(h/n)\;d(nV).
\end{equation}
\noindent We see that in such a system the thermal energy is received due to the change of the number of particles. In cosmology, this change may be considered as a transformation of energy from gravitational field to the matter.\\
\indent In the context of an open system, Eq.(5) can be rewritten as
\begin{equation}
 d(\rho_m V)=-(p_m+p_c)dV,
\end{equation}
\noindent where
\begin{equation}
 p_c=-(h/n)(dN/dV).
\end{equation}
\noindent Equation (6) suggests that the creation of matter in an open thermodynamic system corresponds to a supplementary pressure $p_{c}$, which must be considered as a part of the cosmological pressure entering into the Einstein field equations (decaying of matter leads to a positive decay pressure) and is equivalent to adding the term $p_c$ given by (7) to the thermodynamic pressure $p_m$. It is to be noted that $p_c$ is negative or zero depending on the presence or absence of particle creation. \\
\indent Since the increment in entropy for an adiabatic system is only caused by creation of matter, the entropy is an extensive property of the system. In present scenario, $S$ is proportional to the number of particles included in the system. Therefore, the entropy change $dS$ for open systems from (2) and (5) becomes
\begin{equation}
 TdS=(h/n)\;d(nV)-\mu\; d(nV)=(TS/N)dN \Rightarrow \frac{dS}{S}=\frac{dN}{N},
\end{equation}
\noindent where $\mu$ is the chemical potential given by $\mu =(h-Ts)/n$, $s=S/V$. Since the second law of thermodynamics is a fundamental law in physics, the presence or absence of particle creation can not affect it. This law basically requires $dS\geq0$, consequently,  Eq. (8) gives
\begin{equation}
 dN\geq0.
 \end{equation}
\noindent The above inequality implies that the space-time can produce matter, whereas the reverse process is thermodynamically not admissible.\\
\indent The basic idea of this entire formulation is to modify the usual energy momentum conservation law in an open thermodynamical system, which leads to the explicit use of a balance equation for the number density of the particles created, in addition to Einstein's field equations. \\
\indent The particle flux vector is given by
\begin{equation}
N^{\alpha}=nu^{\alpha},
\end{equation}
\noindent and $N^{\alpha}$ is assumed to satisfy the balance equation$^{44,79}$
\begin{equation}
N^{\alpha}_{\;\;;\alpha}=\psi,
\end{equation}
\noindent where the function $\psi$ denotes a source term of particle creation which is positive or negative depending on whether there is production or annihilation of particles. In standard cosmology $\psi$ is usually assumed to be zero. \\

\noindent\textbf{3. A brief review of Modified $f(R,T)$ gravity theory}\\

\noindent The $f(R,T)$ theory is a modified theory of gravity, in which the Einstein-Hilbert Lagrangian is modified by replacing Ricci scalar curvature $R$ by an arbitrary function of $R$ and trace $T$ of energy-momentum tensor, i.e., $f(R,T)$. So the gravitational action for $f(R,T)$ modified theory of gravity$^{58}$ in the units $ G=1=c$ takes the following form.
\begin{equation}
S=\frac{1}{8\pi}\int d^{4}x\sqrt{-g}\left[\frac{f(R,T)}{2}+\mathcal{L}_{m}\right],
\end{equation}
\noindent where $g$ is the determinant of the metric tensor $g_{\mu\nu}$ and $\mathcal{L}_{m}$ corresponds to matter Lagrangian.  The energy-momentum tensor $T_{\mu\nu}$, defined by fluid Lagrangian density is given by
 \begin{equation}
  T_{\mu\nu}=-\frac{2}{\sqrt{-g}}\frac{\delta(\sqrt{-g}\;\mathcal{L}_m)}{\delta g^{\mu\nu}},
\end{equation}
\noindent and its trace, $T=g^{\mu\nu}T_{\mu\nu}$. By assuming that the matter Lagrangian density $\mathcal{L}_m$ depends only on the metric tensor components $g_{\mu\nu}$, not on its derivatives, we obtain
\begin{equation}
  T_{\mu\nu}=g_{\mu\nu}\mathcal{L}_m-2\frac{\partial \mathcal{L}_m}{\partial g^{\mu\nu}}.
\end{equation}
\noindent The equations of motion by varying the action (12) with respect to metric tensor become
\begin{equation}
f_{R}(R,T)R_{\mu\nu}-\frac{1}{2}f(R,T)g_{\mu\nu}+(g_{\mu\nu}\square-\nabla_\mu\nabla_\nu)f_{R}(R,T)=
8\pi T_{\mu\nu}-f_{T}(R,T)T_{\mu\nu}-f_{T}(R,T)\circleddash_{\mu\nu},
\end{equation}
\noindent where $f_R$ and $f_T$ denote the derivatives of $f(R,T)$ with respect to $R$ and $T$, respectively. Here, $\nabla_\mu$ is covariant derivative and $\square\equiv\nabla_\mu\nabla^\mu$ is the d'Alembert operator and $\circleddash_{\mu\nu}$ is defined by
\begin{equation}
  \circleddash_{\mu\nu}\equiv g^{\alpha\beta}\frac{\delta T_{\alpha\beta}}{\delta g^{\mu\nu}}.
\end{equation}
\noindent Using (14) into (16), we get
\begin{equation}
\circleddash_{\mu\nu}=-2T_{\mu\nu}+g_{\mu\nu}\mathcal{L}_m-2g^{\alpha\beta}\frac{\partial^2\mathcal{L}_m}{\partial g^{\mu\nu}\partial g^{\alpha\beta}}.
\end{equation}
\noindent Since the field equations of $f(R,T)$ gravity depend on $ \circleddash_{\mu\nu}$, i.e.,  on the physical nature of the matter, several models corresponding to different form of $f(R, T)$ would generate depending on the nature of the matter source. The gravitational field equations can be recast in such a form that the higher order corrections coming both from the geometry , and from matter -geometry coupling , provide a stress-energy tensor of geometrical and matter origin, describing an effective source term on right hand side of the standard field equations.  The equations of $f(R,T)$ gravity are much more complicated with respect to the ones of General Relativity even for FRW metric. For this reason many possible form of $f(R,T)$, for example, $f(R,T)= R+2f(T)$,  $f(R,T)$= $\mu f_1(R)+\nu f_2(T)$, where $f_1(R)$ and $f_2(T)$ are arbitrary functions of $R$ and $T$, and $\mu$ and $\nu$ are real constants, respectively $^{58-62}$,  and $f(R,T)=R\;f(T)$ $^{68}$, etc., may be considered to solve the field equations.  We however do not consider the general case, but restrict ourselves to the following form
\begin{equation}
f(R,T)=R+2f(T),
\end{equation}
\noindent where $f(T)$ is an arbitrary function of the trace of energy-momentum tensor of matter. The term $2f(T)$ in the gravitational action modified the gravitational interaction between matter and curvature.\\
\indent Using (18), one can re-write the gravitational field equations defined  in (15) as
\begin{equation}
R_{\mu\nu}-\frac{1}{2}R g_{\mu\nu}=8 \pi T_{\mu\nu}-2 (T_{\mu\nu}+\circleddash_{\mu\nu})f'(T)+f(T) g_{\mu\nu}.
\end{equation}
\noindent Here, a prime stands for derivative of $f(T)$ with respect to $T$. \\
\indent The main issue now arises on the matter contents of the universe through the energy momentum tensor and consequently on the matter Lagrangian $\mathcal{L}_m$ and the trace of the energy momentum tensor. We assume the universe is filled with a perfect fluid which is incorporated in the next section.\\

 \noindent \textbf{4. Model and field equations}\\

\indent We consider a homogenous and isotropic universe with spatially flat geometry described by flat Friedmann-Robertson-Walker (FRW) metric
\begin{equation}
ds^{2} =dt^{2}-a^2(t)(dx^2+dy^2+dz^2),
\end{equation}
where $a(t)$ is the scale factor, which is a function of cosmic time $t$ only.\\
\indent In the formalism of particle creation, the second law of thermodynamics leads naturally to a modification of the energy momentum tensor with an additional creation pressure depends on the rate of creation of particles. In the presence of particle creation, the energy momentum tensor of perfect fluid is given by
\begin{equation}
  T_{\mu\nu}=(\rho_{m}+p_m+p_c)u_\mu u_\nu-(p_m+p_c)g_{\mu\nu},
\end{equation}
\noindent where $u_\mu$ is the four velocity of the fluid such that $u_\mu u^\nu=1$, and in comoving coordinates $u^\mu=\delta_0^\mu$.\\
\indent The trace of energy momentum tensor (21), gives
\begin{equation}
  T=\rho_m-3(p_m+p_c).
\end{equation}
\noindent We treat the scalar invariant $\mathcal{L}_m$ as the effective pressure of the perfect fluid matter and pressure originated by creation of particles. Therefore, the matter Lagrangian may be assumed as $\mathcal{L}_m=-(p_m+p_c)$, therefore, Eq. (17) becomes
\begin{equation}
\circleddash_{\mu\nu}= -2T_{\mu\nu}-(p_m+p_c)g_{\mu\nu}.
\end{equation}
\noindent In view of (23), the field equations (19) can be rewritten as
\begin{equation}
  R_{\mu\nu}-\frac{1}{2}g_{\mu\nu}R=8\pi T_{\mu\nu}+2\left[T_{\mu\nu}+g_{\mu\nu}(p_m+p_c)\right]f'(T)+g_{\mu\nu}f(T).
\end{equation}
\noindent The field equations (24) for a fluid endowed with matter creation (21) in the background of metric (20), yield
\begin{eqnarray}
  3H^2&=&8\pi\rho_m+2(\rho_m+p_m+p_c)f'(T)+ f(T),\\
 2\dot H+3H^2&=&-8\pi (p_m+p_c) + f(T),
\end{eqnarray}
\noindent where $H(t)=\dot a/a$ is the Hubble parameter. A dot denotes derivative with respect to cosmic time $t$.\\
\indent For adiabatic matter creation, the pressure $p_{c}$ in Eq. (7) takes the following form$^{44}$
\begin{equation}
p_{c}=-\frac{(\rho_{m}+p_{m})}{3nH}\;\psi.
\end{equation}
\noindent For a complete formulation of the problem, we consider one more relation between the particle number $n$ and $V = a^{3}$, describing the dynamics of $n$ as a result of matter creation (decay) processes. This relation is given by Eq. (11), which for the metric (20) takes the form
\begin{equation}
\dot{n}+3nH=\psi,
\end{equation}

\noindent\textbf{5. Solution of field equations }\\

\noindent The field equations (25)-(28) have seven unknowns, namely, $a$, $\rho_m$, $p_m$, $p_c$, $n$, $f(T)$ and $\psi(t)$. We have only three equations plus the constraint (27). Therefore, one needs three more relations in order to construct a definite cosmological scenario. \\
\indent In first choice, we consider a particular function given by$^{58}$
\begin{equation}
  f(T)=\lambda T, \;\;\;\;(\lambda \;\;\; \text{is a constant}).
\end{equation}
\noindent With this assumption, the field equations (25) and (26) yield
\begin{eqnarray}
  3H^2&=&(8\pi+3\lambda)\rho_m-\lambda(p_m+p_c),\\
 2\dot H+3H^2&=&-(8\pi+3\lambda)(p_m+p_c)+\lambda\rho_m.
\end{eqnarray}
\noindent In order to obtain the exact solution of the field equations, we assume two more additional relations: the equation of state and the matter creation rate. In the cosmological domain, the former is usually given by the gamma-law
\begin{equation}
p_m=(\gamma-1)\rho_m,
\end{equation}
\noindent where $\gamma$ is a constant lies in the interval [0, 2] and known as the EoS parameter of the perfect fluid. \\
\indent Using (32) into (30) and (31), and simplifying, we get a single evolution equation for $H$:
\begin{equation}
2\dot{H}+(8\pi+2\lambda)(\gamma\rho_m+p_{c})=0.
\end{equation}
\noindent We confine our attention to the simple phenomenological expression for the matter creation rate$^{21}$\\
\begin{equation}
\psi(t)=3\beta n H,
\end{equation}
\noindent where the parameter $\beta$ lies in the interval $(0,1)$ and is assumed to be a constant.\\
\indent Using (32) and (34) into (27), we have
\begin{equation}
p_c=-\beta\gamma\rho_m,
\end{equation}
\noindent Putting (32) and (35) into (30), we obtain
\begin{equation}
\rho_{m}= \frac{3H^2}{8\pi +4\lambda-\gamma \lambda(1-\beta)}.
\end{equation}
\noindent Substituting (35) and (36) into (33) gives
\begin{equation}
\dot H+\frac{3}{2}\frac{\gamma(8\pi + 2\lambda)(1-\beta)}{\left[8\pi+4\lambda-\gamma \lambda(1-\beta) \right]}\;H^2=0.
\end{equation}
\noindent The solution of (37) for $\gamma\neq0$ and for all values of $\lambda$ and $\beta$ is given by
\begin{equation}
H(t)=\left(C+\frac{3}{2}\frac{\gamma(8\pi + 2\lambda)(1-\beta)}{\left[8\pi+4\lambda-(1-\beta)\gamma \lambda \right]}\;t\right)^{-1},
\end{equation}
where $C$ is an integration constant. For $\gamma =0$, the well known de-Sitter scale factor $a(t)=a_{0}\;e^{H_{0}t}$ is obtained.\\
\indent From Eq. (38) we find the following expression for the scale factor
\begin{equation}
a(t)=D\left(C+\frac{3}{2}A \gamma\;t \right)^{\frac{2}{3A\gamma}},
\end{equation}
\noindent where $D$ is a new  integration constant and $A=(8\pi + 2\lambda)(1-\beta)/8\pi+4\lambda-\gamma\lambda(1-\beta)$.\\
\indent The above scale factor may be rewritten as
\begin{equation}
a(t)=a_{0}\left(1+\frac{3}{2}A \gamma H_{0}(t-t_{0}) \right)^{\frac{2}{3 A\gamma}},
\end{equation}
\noindent where  $H=H_{0}>0$ at $t=t_{0}$. The subscript `$0$' refers to the present values of parameters. Since $0\leq\gamma\leq2$, we must have $A>0$ for expansion of the universe. Also, $A>0$ implies $\lambda>0$ as $0\leq\beta<1$.\\
\indent If $\gamma < 0$ then we have a Big Rip singularity at a finite value of cosmic time $t_{br}=t-t_{0}=-2/3H_{0}A\gamma$. Thus, we have a phantom cosmology for $\gamma<0$. If one choose $t_{0}=2H^{-1}_{0}/3A\gamma$ then (40) takes the familiar form of power-law expansion of the universe, i.e.,
\begin{equation}
a(t)=a_{0}\left(\frac{3}{2}A \gamma H_{0}\;t \right)^{\frac{2}{3 A\gamma}}.
 \end{equation}
\noindent If $\lambda=0=\beta$, (39) and (41) reduce to the well-known expressions of power-law expansion of scale factor for a flat FRW model in GR.\\
\indent By use of (40) one may express the energy density of matter, particle creation pressure and the particle number density as functions of the scale factor $a$. These parameters respectively have the following forms
 \begin{eqnarray}
 \rho_m&=&\rho_0 \left(\frac{a_{0}}{a}\right)^{3A\gamma},\\
 p_{c}&=&-\beta \gamma \rho_{0}\left(\frac{a_{0}}{a}\right)^{3A\gamma},\\
 n&=&n_{0}\left(\frac{a_{0}}{a}\right)^{3(1-\beta)},
 \end{eqnarray}
\noindent where $\rho_{0}= 3H^{2}_{0}/[8\pi+4\lambda-\gamma \lambda(1-\beta)]$ is the present value of energy density. Here, $n_0$ is the present value of particle number density for any values of $\beta$. The above results show that the transition from one phase to other phase, in the course of expansion, happens exactly as in the standard cosmological model.\\
\indent The number of particles $N$ in a given volume $V$ is given by
\begin{equation}
 N=N_0\left(\frac{a}{a_0}\right)^{3\beta},
 \end{equation}
 \noindent which shows that $N$ increases with time. If $\beta = 0$, N would remain constant throughout the
evolution of the universe and we would recover the standard FRW model of the universe in $f(R,T)$ theory. Again, from (8) $S=S_{0}(N/N_0)$, the entropy in terms of scale factor is
 \begin{equation}
 S=S_0\left(\frac{a}{a_0}\right)^{3\beta}.
 \end{equation}
\noindent The deceleration parameter defined as $q=-a\ddot a/\dot a^2$, gives
\begin{equation}
q= -1+\frac{3\gamma A}{2}=\left[\frac{3\gamma}{2}\frac{(8\pi+2\lambda)(1-\beta)}{[8\pi+4\lambda-(1-\beta)\gamma \lambda]}-1\right].
\end{equation}
\noindent which shows that $q$ is independent of cosmic time $t$. Therefore, q may be positive or negative for a given set of values of $\beta$ and $\lambda$. We know that the model accelerates for $q<0$, therefore, the value of $A$ must be $0<A<2/3\gamma$ for an accelerated universe.\\
\indent As expected, the above solutions reduce to the standard FRW model of GR for $\beta=0$ and $\lambda=0$ and for all values of $\gamma$. In what follows, we study the role of $f(R,T)$ gravity and particle creation in early and late time evolution of the universe.\\

\noindent \textbf{ Case I: $\gamma=\frac{2}{3}$}\\

\noindent Fig. 1(a) plots the scale factor as a function of time for $\gamma=2/3$ and some selected values of $\lambda$ and $\beta$. We observe that if $\beta=0$, $q<0$ for all $\lambda>0$, therefore, we find that the universe accelerates in $f(R,T)$ gravity without particle creation. Similarly, if $\lambda=0$, i.e., in the absence of $f(T)$, $q=-\beta<0$ for any values of $\beta>0$. Thus, the acceleration occurs due to particle creation. The rate of expansion increases more rapidly for non-zero values of $\beta$ and $\lambda$. It is to be noted that if $\lambda=0=\beta$ then the the marginal inflationary phase of GR is recovered, i.e., $a\sim t$ and $q=0$.\\
{\begin{center}
\begin{tabular}{cc}
\begin{minipage}{200pt}
\frame{\includegraphics[width=200pt]{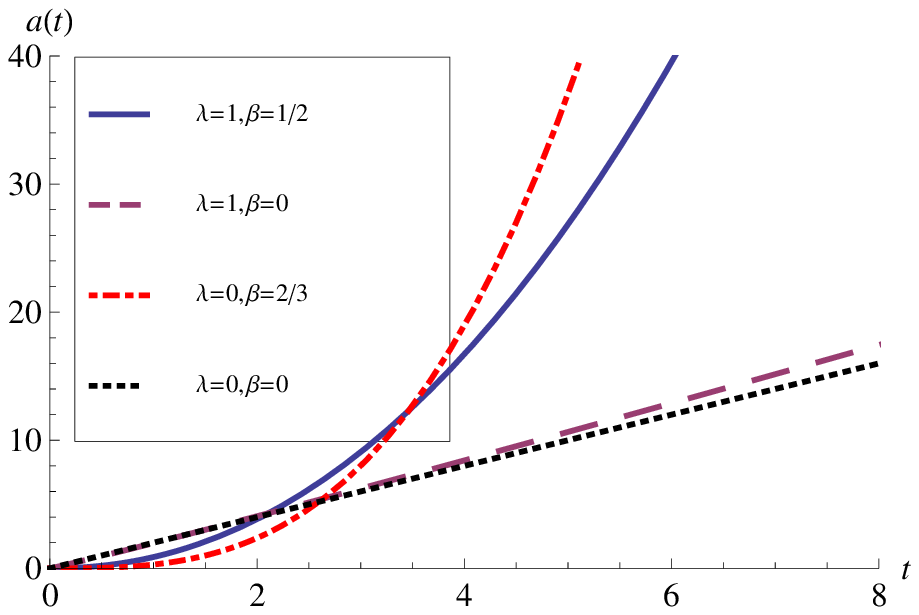}}
\center{\footnotesize Fig. 1(a). Scale factor as a function of time for $\gamma=\frac{2}{3}$ and some selected values of $\lambda$ and $\beta$. }
\end{minipage}&\begin{minipage}{200pt}
\frame{\includegraphics[width=200pt]{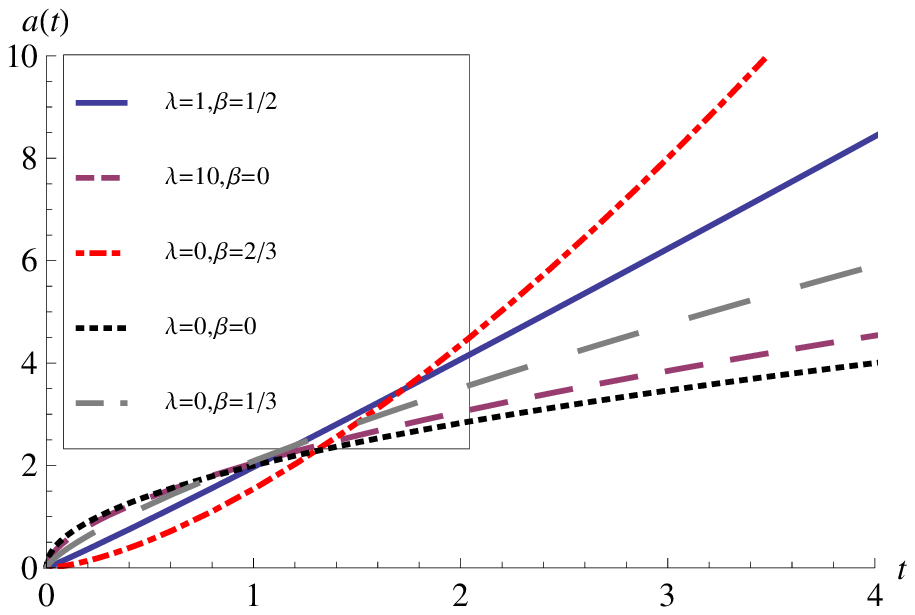}}
\center{\footnotesize Fig.1(b). Scale factor as a function of time for $\gamma=\frac{4}{3}$ and some selected values of $\lambda$ and $\beta$.}
\end{minipage}
\end{tabular}
\end{center}}

\noindent \textbf{Case II:  $\gamma=\frac{4}{3}$}\\

\noindent In this case, if $\beta=0$ and $\lambda>0$, we have $q>0$. This shows that the universe decelerates in the absence of particle creation. If $\lambda=0$ then $q\geq0$ for $0<\beta\leq1/2$, and $q<0$ for $1/2<\beta<1$. Therefore, the universe decelerates or accelerates depending on the rate of particle creation. However, if $\lambda\neq0$ and $\beta\neq0$, the deceleration or acceleration of the universe depend on the following constrains, respectively:
\begin{equation}
 0<\beta \leq \frac{1}{4},\;\lambda >0\;\; or\;\; \frac{1}{4}<\beta <\frac{1}{2},\;0<\lambda <\frac{6 \pi -12 \pi  \beta }{-1+4 \beta },
\end{equation}
\begin{equation}
\frac{1}{4}<\beta \leq \frac{1}{2},\;\lambda >\frac{6 \pi -12 \pi  \beta }{-1+4 \beta }\;\;or\;\;\frac{1}{2}<\beta <1,\;\lambda >0.
\end{equation}
\noindent The behavior of scale factor versus time is shown in Fig. 1(b) for some selected values of $\lambda$ and $\beta$. The figure shows that the universe accelerates fast due to higher particle creation rate. For $\lambda=0=\beta$, $a\sim t^{1/2}$ and $q=1$, which is the radiation phase in GR.\\

\noindent \textbf{ Case III:  $\gamma=1$}\\

\noindent In this case, the universe expands with decelerated rate as $q>0$ for $\beta=0$ and $\lambda>0$. Fig. 1(c) plots graph between scale factor versus time for some selected values of $\lambda$ and $\beta$. For $\lambda=0$, we have $q>0$ for $0<\beta<1/3$, and $q<0$ for $1/3<\beta<1$. The critical case ($\beta=1/3$, $q=0$), describes a coasting cosmology. For $\lambda\neq0$ and $\beta\neq0$, the model decelerates or accelerates under the following constraints:
\begin{equation}
0<\beta <\frac{1}{3},\;0<\lambda <\frac{\pi -3 \pi  \beta }{\beta },
\end{equation}
\begin{equation}
0<\beta \leq \frac{1}{3},\;\lambda >\frac{\pi -3 \pi  \beta }{\beta }\;\;or\;\;\frac{1}{3}<\beta <1,\;\lambda >0,
\end{equation}
\noindent respectively. As expected, for $\lambda=0=\beta$, we have $a\sim t^{2/3}$ and $q=1/2$, i.e., the model reduces to standard matter-dominated era of GR.\\
{\begin{center}
\begin{tabular}{cc}
\begin{minipage}{200pt}
\frame{\includegraphics[width=200pt]{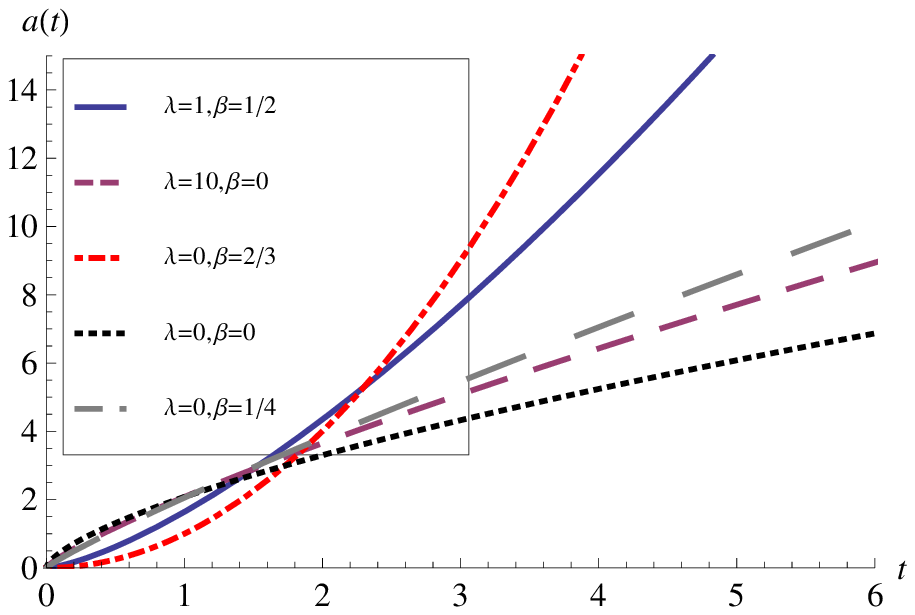}}
\center{\footnotesize Fig. 1(c). Scale factor as a function of time for $\gamma=1$ and some selected values of $\lambda$ and $\beta$. }
\end{minipage}&\begin{minipage}{200pt}
\frame{\includegraphics[width=200pt]{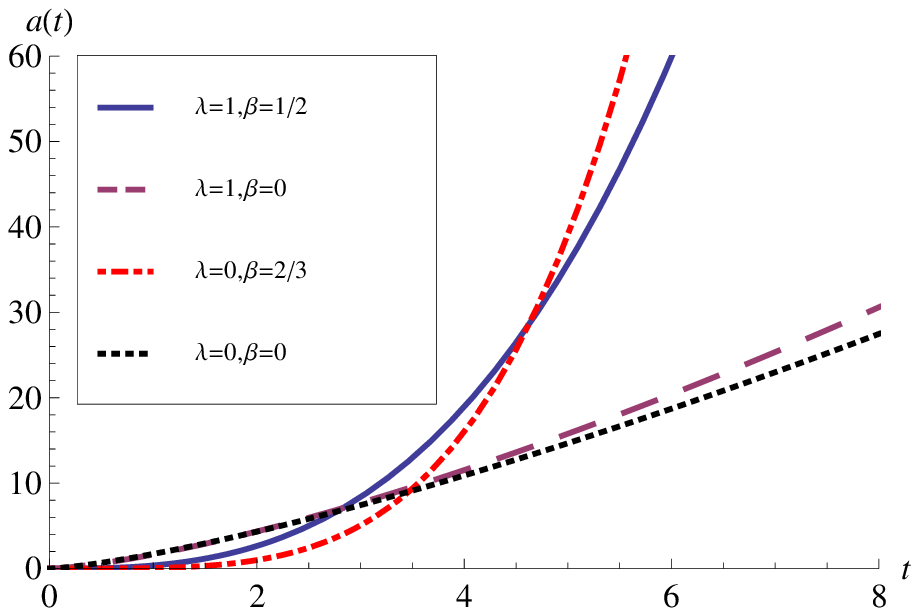}}
\center{\footnotesize Fig. 1(d). Scale factor as a function of time for $\gamma=\frac{1}{2}$ and some selected values of $\lambda$ and $\beta$.}
\end{minipage}
\end{tabular}
\end{center}}

\noindent \textbf{Case IV:  $\gamma=\frac{1}{2}$} \\

\noindent In this case, if $\lambda=0=\beta$, $a\sim t^{4/3}$ and $q=-1/4$, which corresponds to the present accelerated phase of the universe of the standard FRW universe in GR. Since the universe accelerates even in absence of both $f(T)$ and particle creation, therefore, the contribution of $f(R,T)$ gravity or particle creation just enhance the rate of acceleration of the universe. Fig. 1(d) plots the dynamics of scale factor versus $t$, which is similar to case I.\\

\noindent\textbf{6. Kinematic tests }\\

\indent Now, we derive some kinematical relations of the  model as proposed in the preceding sections.\\\\

\noindent\textbf{6.1 The density parameter}\\

\indent The density parameter, defined as $\Omega_m=\rho_m/\rho_c$, where $\rho_c=3H^{2}/8\pi $, is given by
\begin{equation}
\Omega_m=\frac{8\pi}{8\pi+4\lambda-(1-\beta)\gamma \lambda}.
\end{equation}
\noindent Therefore, it is clear that $\Omega_m<1$ for all values of $0\leq\gamma\leq2$, $0<\beta<1$ and $\lambda>0$. Hence the universe is negatively curved. In absence of both $\lambda$ and $\beta$, we have $\Omega_m=1$ for all $\gamma$, i.e., the flat model of GR is recovered.\\

\noindent\textbf{6.2 Lookback time-redshift }\\

\noindent The lookback time $\Delta t=t_0-t(z)$, is the difference between the age of the universe at the present time $z=0$ and the age of the universe when a particular right ray at redshift $z$ was emitted. \\
\indent For a given redshift $z$, the scale factor $a(t_z)$ is related to $a_0$ by
\begin{equation}
 a(t_z)= a_0(1+z)^{-1}.
\end{equation}
\noindent From (41) and (53), the cosmic time in terms of redshift is given by
\begin{equation}
  t(z)=\frac{2H_0^{-1}}{3\gamma A}(1+z)^{-\frac{3\gamma A}{2}},
\end{equation}
\noindent where $H_0$ is expressed in the usual practical observational units of $km\;s^{-1}\;Mpc^{-1}$ and its value is believing to be somewhere between 50-80 $km\;s^{-1}\;Mpc^{-1}$. However, $H_0$ is dimensionally similar to the reciprocal of time. The reciprocal of Hubble constant is called the Hubble time $t_H:t_H=H_0^{-1}$, where $t_H$ is expressed in $s$ and $H_0$ in $s^{-1}$. If $H_0$ is expressed in $km\;s^{-1}\;Mpc^{-1}$ and $t_H$ in $gigayears\; (1\; gr=1\;milion\;years=10^9\;years)$ then $t_H=977.8/H_0$.\\
\indent Therefore, from (54) we have
\begin{equation}
  t_0-t(z)=\frac{2H_0^{-1}}{3\gamma A}\left[1-(1+z)^{-\frac{3\gamma A}{2}}\right].
\end{equation}
\noindent Fig. 2 plots lookback time versus redshift for $\gamma=1$ and some selected values of $\lambda$ and $\beta$. All models coincide for lower redshift since they follow the same behavior. The graph shows that the lookback time increases for higher values of $\beta$. Thus, the universes with larger matter creation rate are older.\\
\indent For small values of redshift, (55) becomes
\begin{equation}
  H_0\left(t_0-t(z)\right)=z-\left(1+\frac{3\gamma A}{2}\right)z^2+\cdots.
\end{equation}
\noindent Taking $\lim\;{z\to\infty}$ in (55), the present age of the universe is
\begin{equation}
  t_0=\frac{2H_0^{-1}}{3\gamma A}=\frac{H_0^{-1}}{1+q}.
\end{equation}
\noindent Thus, the age of the universe depends on both parameters $\beta$ and $\lambda$.
{\begin{center}
\begin{tabular}{cc}
\begin{minipage}{200pt}
\frame{\includegraphics[width=200pt]{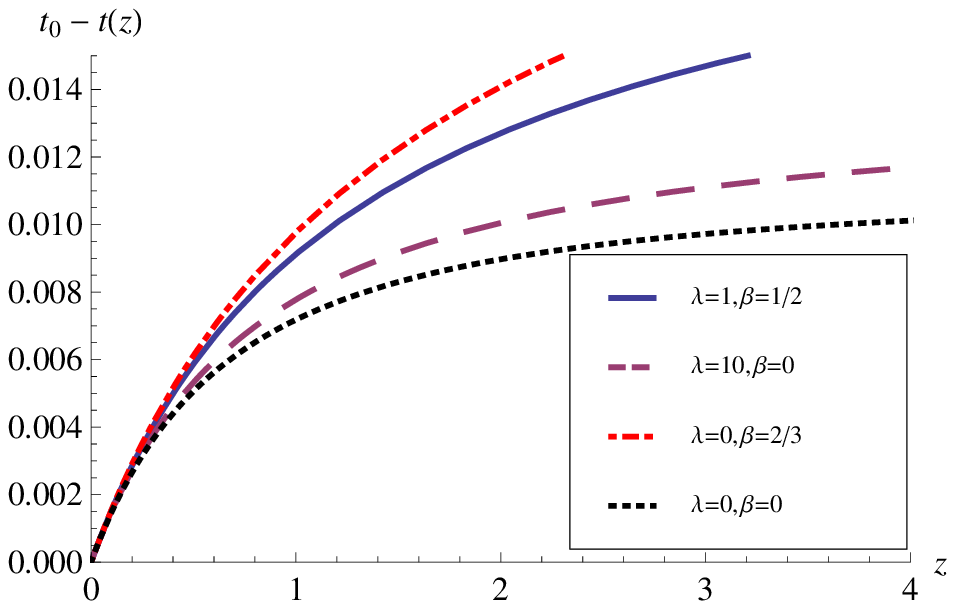}}
\center{\footnotesize Fig. 2. Lookback time versus redshift for $\gamma=1$, $H_0=60$ and some selected values of  $\lambda$ and $\beta$.}
\end{minipage}&\begin{minipage}{200pt}
\frame{\includegraphics[width=200pt]{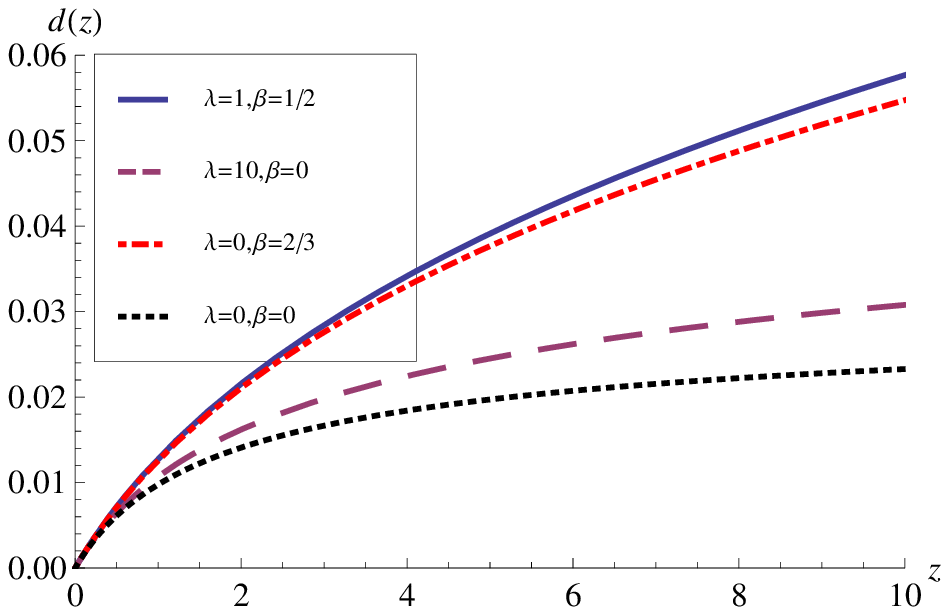}}
\center{\footnotesize Fig. 3. Proper distance versus redshift for $\gamma=1$, $H_0=60$ and some selected values of  $\lambda$ and $\beta$.}
\end{minipage}
\end{tabular}
\end{center}}

\noindent\textbf{6.3 Proper distance-redshift }\\

\noindent The proper distance between the source and observer is defined as $d(z)=a_0r(z)$, where $r(z)$ is the radial distance of the object at light emission in term of redshift given by
\begin{equation}
  r(z)=\int_t^{t_0}\frac{dt}{a(t)}=\frac{H_0^{-1}}{a_0\left(\frac{3\gamma A}{2}-1\right)}
  \left[1-(1+z)^{-\left(\frac{3\gamma A}{2}-1\right)}\right].
\end{equation}
\noindent Consequently, the proper distance becomes
\begin{equation}
  d(z)=\frac{H_0^{-1}}{\left(\frac{3\gamma A}{2}-1\right)}
  \left[1-(1+z)^{-\left(\frac{3\gamma A}{2}-1\right)}\right].
\end{equation}
\noindent The proper distance as a function of redshift for some selected values of $\beta$ and $\lambda$ are displayed in Fig. 3. We observe that the $f(T)$ contribution in $f(R,T)$ and particle creation gives rise to proper distance. \\
\indent Equation (59) can be rewritten as
\begin{equation}
  H_0d(z)=z-\frac{3\gamma A}{4}z^2+\cdots .
\end{equation}
\noindent From  (59), it is observed that the distance $dz$ is maximum at $z\to\infty$. Hence,
\begin{equation}
  H_0d(z\to\infty)=\frac{1}{\frac{3\gamma A}{2}-1}=\frac{1}{q}.
\end{equation}\\

\noindent\textbf{6.4 Luminosity distance-redshift }\\

\noindent The best-known way to trace the evolution of the universe observationally is to look into the redshift-luminosity distance relation. The luminosity distance $d_l$ is defined by the relation $d_l^2=\frac{l}{4\pi L}$, where $l$ is the luminosity of the object and $L$ is the measured flux from the object. In standard FRW cosmology it is defined in terms of redshift as
\begin{equation}
  d_l=a_0(1+z)r(z)=(1+z)d(z).
\end{equation}
\noindent From (59) and (62), we get
\begin{equation}
  d_lH_0=\frac{1}{\left(\frac{3\gamma A}{2}-1\right)}
  \left[(1+z)-(1+z)^{-\left(\frac{3\gamma A}{2}-2\right)}\right].
\end{equation}
\noindent The graph between Luminosity distance and redshift for some selected values of $\beta$ and $\lambda$ is plotted in Fig. 4. One may observe that the luminosity distance corresponding to any specific value of redshift rises due to $f(R,T)$ gravity and particle creation.\\
\indent Expanding (63) for small z, we find
\begin{equation}
  H_0d_l=z-\frac{1}{2}\left(\frac{3 \gamma A}{2}-2\right)z^2+\cdots.
\end{equation}
 As expected, we find the same behavior for different models at $z\ll1$ and the possible difference in behaviors for different models come at large redshift $(z\gg1)$. In Fig. 4 we observe that all curves start off with the linear Hubble law $(z = d_l H_0)$ for small $z$, but then, only the curve for $q = 1$, i.e., $\beta = 0 =\lambda$ stays linear all the way. We also note that for the small redshift the luminosity distance is larger for lower values of q.
{\begin{center}
\begin{tabular}{cc}
\begin{minipage}{200pt}
\frame{\includegraphics[width=200pt]{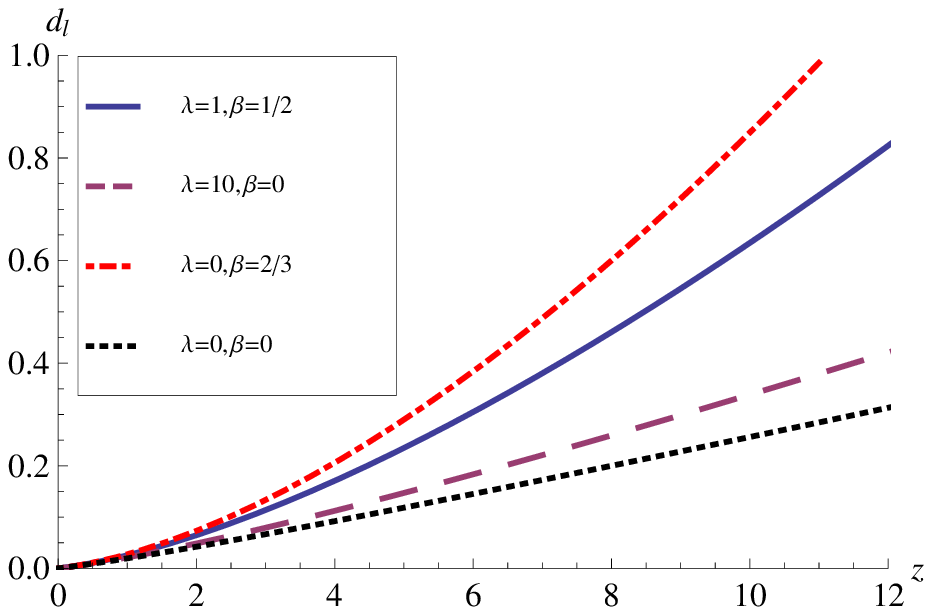}}
\center{\footnotesize Fig. 4. Luminosity distance versus redshift for $\gamma=1$, $H_0=60$ and some selected values of  $\lambda$ and $\beta$.}
\end{minipage}&\begin{minipage}{200pt}
\frame{\includegraphics[width=200pt]{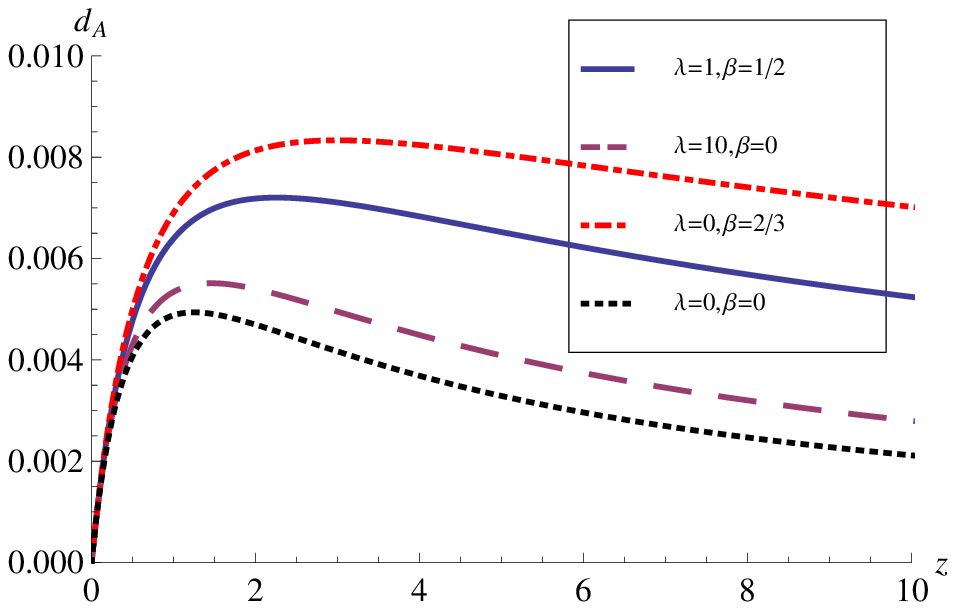}}
\center{\footnotesize Fig. 5. Angular diameter distance versus redshift for $\gamma=1$, $H_0=60$ and some selected values of  $\lambda$ and $\beta$.}
\end{minipage}
\end{tabular}
\end{center}}

\noindent\textbf{6.5 Angular diameter distance-redshift }\\

\noindent The angular diameter distance $d_A$ is the ratio of physical transverse size of an object to its angular size (in radians). In terms of $z$, it is given by
\begin{equation}
  d_A=\frac{d(z)}{1+z}=\frac{d_l}{(1+z)^2}.
\end{equation}
\noindent Using (59), we have
\begin{equation}
 H_0d_A=\frac{1}{\left(\frac{3\gamma A}{2}-1\right)}
  \left[(1+z)^{-1}-(1+z)^{-\frac{3\gamma A}{2}}\right].
\end{equation}
\noindent In Fig. 5 we plot the angular diameter distance versus redshift for some selected values of $\beta$ and $\lambda$. The graph shows that the $f(R,T)$ and particle creation enhance the angular distance. The angular diameter distance initially increases with increasing $z$ and eventually begins to decrease.\\
\indent On expanding (66), we get
\begin{equation}
 H_0d_A=z+\left[1-\frac{\left(\frac{3\gamma A}{2}+1\right)\left(\frac{3\gamma A}{2}+2\right)}{2\left(\frac{3\gamma A}{2}-1\right)}\right]z^2+\cdots.
\end{equation}

\noindent \textbf{7 Conclusion}\\

\noindent In this paper, we have studied a flat FRW cosmological model described by an open thermodynamic system including particle creation at the expense of gravitational field in $f(R,T)$ theory of gravity. We have obtained exact solutions for the scale factor and various physical quantities by assuming a suitable form of $f(R,T)=R+2f(T)$ and ``gamma-law" equation of state. The model exhibits non-singular power-law expansion of the universe for $0\leq\gamma\leq2$. The model shows Big Rip singularity for $\gamma<0$. The dynamics of the scale factor and other physical quantities have been examined through some graphical representations in various phases of evolution of the universe with some selected values of $\lambda$ and $\beta$.\\
\indent It has been observed that the scale factor always increases with decelerated and accelerated rate depending upon the contribution of particle creation and the parameter $\lambda$. It has been noticed that the universe expands with slow rate in early time but fast during late time. The energy density and effective pressure always decrease with time and both tend to zero in late time for $0\leq\gamma\leq2$. The number of particles increase with time in all phases. The number of particles in the absence of particle production remain constant throughout the evolution of the universe, which is quit obvious. The deceleration parameter has been found as a constant, which exhibits both decelerated and accelerated universe under some constraints on different parameters. The density parameter shows that the model becomes open in the presence of particle creation in $f(R,T)$ theory.\\
\indent We have also discussed some observational consequences of the model through some kinematics tests such as lookback time, proper distance, luminosity distance and angular diameter distance with respect to redshift. The results for the cosmological tests are found to be compatible with the present observations. These tests are found to be depending on $\lambda$ and $\beta$. The universe with particle creation is always older than the usual FRW model in absence of particle creation. The model of Lima et al.$^{21}$ may be recovered for $\lambda=0$.\\
\indent In summary, we have studied a cosmological model with particle creation in $f(R,T)$ gravity theory to understand the current acceleration expansion of the universe. We have found that the negative pressure due to the matter creation may play the role of dark energy and derive the accelerating expansion of the universe in $f(R,T)$ theory. We may expect that the process of particle creation is also an ingredient which accounts this unexpected observational results. The changes introduced by the particle creation process, which is quantified by the parameter $\beta$, provide reasonable observational results. The new fact justifying the present work is that we have considered the thermodynamics approach for which particle creation is at the expense of the gravitational field. A general expression relating the energy densities and particle number density as function of scale factor have been established. One may find that the particle creation changes the predictions of standard cosmology, thereby alleviating the problem of reconciling observations with the inflationary scenario. In future work, we plan to constraint cosmological model with matter creation using complimentary astronomical observations.\\

\noindent\textbf{Acknowledgement} VS expresses his sincere thank to University Grants Commission (UGC), India, for providing Senior Research Fellowship (SRF).\\

\noindent \textbf{References}\\
\small{
\begin{description}
%Ia Supenova
\item[]1. A.G. Riess et al., {\it Astrophys. J.} {\bf 116}, 1009 (1998).
\item[]2. S. Perlmutter et al., {\it  Astrophys. J.} {\bf 517}, 565 (1999).
\item[]3. R.G. Vishwakarma, {\it Mon. Not. Roy. Astron. Soc.} {\bf 331}, 776 (2002).
%CMB
\item[]4. C.B. Netterfield et al., {\it  Astrophys. J.} {\bf 571}, 604 (2002).
%LSS
\item[]5. D.N. Spergel et al., {\it  Astro J. Suppl.} {\bf 148}, 175 (2003)  [astro-ph/0302209].
%DE
\item[]6. V. Sahni, {\it  Lec. Notes Phys.} {\bf 653}, 141 (2004)  [astro-ph/0403324].
\item[]7. J. Frieman et al., {\it  Ann. Rev. Astron. Astrophys.} {\bf46}, 385 (2008)  [astro-ph/0803.0982].
\item[]8. R.R. Caldwell and M. Kamionkowski,  {\it  Ann. Rev. Nucl. Part. Sci.} {\bf59},  397 (2008)  [astro-ph/0903.0866].
%gravitational effect
\item[]9. L.M. Krauss and  M. Turner, {\it  Gen. Rel. Grav.} {\bf 27}, 1137 (1995).
\item[]10. M.S. Turner and M. White, {\it  Phys. Rev. D} {\bf 56}, R4439 (1997).
\item[]11. T. Chiba, N. Sugiyama and T. Nakamura, {\it Mon. Not. Roy. Astron. Soc.} {\bf 289}, L5 (1997).
%Cosmo Const.
\item[]12. T. Padmanabhan {\it Phys. Rept.} {\bf 380}, 235 (2003)  [hep-th/0212290].
%Quintessence
\item[]13. J. Martin,  {\it Mod. Phys. Lett. A} {\bf 23}, 1252 (2008)  [astro-ph/0803.4076].
%Phantom
\item[]14. U. Alam et al.,  {\it Mon. Not. Roy. Astron. Soc.} {\bf 354}, 275 (2004)
%kessence
\item[]15. T. Chiba et al., {\it Phys. Rev. D} {\bf62}, 023511 (2000).
%tachyon
\item[]16. T. Padmanabhan and T.R. Chaudhury, {\it Phys. Rev. D} {\bf 66}, 081301 (2002).
%chaplygin
\item[]17. M.C. Bento et al., {\it Phys. Rev. D} {\bf66}, 043507 (2002).
%singularity problem
\item[]18. L.R.W. Abramo and J.A.S. Lima, {\it Class. Quant. Grav.} {\bf 13}, 2953 (1996).
%reheating
\item[]19. W. Zimdhal and P. Pav$\acute{o}$n, {\it Mon. Not. Roy. Astron. Soc.} {\bf 266}, 872 (1994).
%age problem
\item[]20. J.A.S. Lima and L.R.W. Abramo, {\it Phys. Lett. A} {\bf 257}, 123 (1999).
%entropy problem
\item[]21. J.A.S. Lima, A.S. Germano and L.R.W.  Abramo, {\it Phys. Rev. D} {\bf 53}, 4287 (1996).
\item[]22. I. Brevik and G. Stokkan, {\it Astrophys. Space. Sci.} {\bf 239},  89 (1996).
%inflation
\item[]23. A.H. Guth, {\it Phys. Rev. D} {\bf 23}, 347 (1981)
\item[]24. A.D. Linde, {\it Phys. Lett.} {\bf 108B}, 389 (1981)
%age confliction
\item[]25. W.L. Freedman, A. Olinto, J. Friemaan and D. Schramm, {\it Proceedings of the 18th texas Symposium on Relativistic Astrophys.} World Scietific (1998) {\bf }.
%study of early universe
\item[]26. C. Brans and R.H. Dicke, {\it Phys. Rev. D} {\bf 124}, 925 (1961).
\item[]27. P.G. Bergmann, {\it Int. J. Theor. Phys.} {\bf 1}, 25 (1968).
\item[]28. K.Nortvedt, {\it Astrophys. J.} {\bf 161}, 1059 (1970).
\item[]29. R.V. Wagoner, {\it Phys. Rev. D} {\bf 1}, 3209 (1970).
%dirac large number
\item[]30. P.A.M. Dirac, {\it Nature} {\bf 139}, 323 (1937).
\item[]31. V. Canuto, P.J. Adoms, S.H. Hsieh and E. Tsiang, {\it Nature} {\bf 261}, 438 (1976).
\item[]32. V. Canuto, P.J. Adoms, S.H. Hsieh and E. Tsiang, {\it Phys. Rev. D} {\bf 16}, 1643 (1977).
%PCP
\item[]33. H. Bondi and T. Gold, {\it Mon. Not. Roy. Astron. Soc.} {\bf 108}, 252 (1948).
\item[]34. F. Hoyle, {\it Mon. Not. Roy. Astron. Soc.} {\bf 109}, 365 (1949).
\item[]35. J.V. Narlikar, {\it Nonstandard cosmologies}, in Vth Brazialian school of cosmology and Gravitation, ed. M. Novello (World Scientific, Singapore) 152 (1987).
\item[]36. E.P. Tryon,  {\it Nature} {\bf 246}, 396 (1973).
\item[]37. P.I. Fomin, {\it Dokl. Akad. Nank SSSR A} {\bf 9}, 831 (1975).
\item[]38. R. Brout, F. Englert and E. Gunzig, {\it Ann. Phys. (NY)} {\bf 115}, 78 (1978).
\item[]39. R. Brout, F. Englert and E. Gunzig, {\it Gen. Rel. Grav.} {\bf 1}, 1 (1979).
\item[]40. R. Brout, F. Englert, F. Fr$\acute{e}$re, J.M. Gunzig, P. Nardone, P. Spindel, C. Tuffin and E. Gunzig, {\it Nucl. Phys. B} {\bf 170}, 228 (1980).
\item[]41. R. Brout, F. Englert and P. Spindel, {\it Phys. Rev. Lett.} {\bf 43}, 417 (1979).
\item[]42. E. Gunzig, J. Gehenian and I. Prigogine, {\it Nature} {\bf 330}, 621 (1983).
\item[]43. I. Prigogine, J. Gohenian, E. Gunzig and P. Nardone, {\it Proc. Natl. Acad. Sci. USA} {\bf 85}, 7428 (1988).
\item[]44. M.O. Calv$\tilde a$o, J.A.S. Lima and I. Waga, {\it Phys. Lett. A} {\bf 162}, 223 (1992).
\item[]45. J.A.S. Lima and A.S. Germano, {\it Phys. Lett. A} {\bf 170}, 373 (1992).
\item[]46. J.A.S. Lima and J.S. Alcaniz, {\it Astron. Astrophys.} {\bf 348}, 1 (1999).
\item[]47. J.S. Alcaniz and J.A.S. Lima,  {\it Astron. Astrophys.} {\bf 349}, 72 (1999).
\item[]48. W. Zimdahl et al., {\it Phys. Rev. D} {\bf 64}, 063501 (2001).
\item[]49. Y. Qiang, T. Jiezhang and Y. Ze-Long, {\it Astrophys. Space Sci.} {\bf 311}, 407 (2007)   [arXiv,astro-ph/0503123].
\item[]50. C.P. Singh and A. Beesham, {\it Astrophys. Space Sci.} {\bf 336}, 469 (2011).
\item[]51. C.P. Singh and A. Beesham, {\it Int. J. Theor. Phys.}  {\bf 51}, 3951 (2012).
\item[]52. C.P. Singh, {\it Astrophys.  Space Sci.} {\bf 338}, 411 (2012).
\item[]53. C.P. Singh, {\it Mod. Phys. Lett. A}  {\bf 27}, 1250070 (2012).
%Review modified gravity
\item[]54. T. Clifton, P.G. Ferreira, A. Padilla and C. Skordis, {\it Phys. Rep.} {\bf 513}, 1 (2012) [arXiv,astro-ph/1106.2476].
%f(R)
\item[]55. S. Nojiri andS.D. Odintsov, {\it Phys. Rev. D} {\bf74}, 086005 (2006).
\item[]56. O. Bertolami, C.G. Bochmer, T. Harko and F.S.N. Lobo, {\it Phys. Rev. D} {\bf75}, 104016 (2007).
\item[]57. T. Harko, {\it Phys. Lett. B} {\bf 669}, 376 (2008).
\item[]58. T. Harko, F.S.N. Lobo, S. Nojiri and S.D. Odintsov, {\it Phys. Rev. D} {\bf 84}, 024020 (2011)   [arXiv,gr-qc/1104.2669].
\item[]59. M. Jamil, D. Momeni, M. Raza and R. Myrzakulov, {\it Eur. Phys. J. C} {\bf72}, 1999 (2012).
\item[]60. M.J.S. Houndjo, {\it Int. J. Mod. Phys. D} {\bf 21}, 1250003 (2012)  [arXiv,astro-ph/1107.3887].
\item[]61. M.J.S. Houndjo and O.F. Piattella, {\it Int. J. Mod. Phys. D} {\bf 2},
1250024 (2012)   [arXiv,gr-qc/1111.4275].
\item[]62. M.J.S. Houndjo, C.E.M. Batista, J.P. Campos and O.F. Piattella, {\it Canadian. J. Phys. } {\bf 91}, 548 (2013)   [arXiv,gr-qc/1203.6084].
\item[]63. M. Sharif and M. Zubair, {\it J. Cosmo. Astropart. Phys.} {\bf 21}, 28 (2012) [arXiv,gr-qc/1204.0848].
\item[]64. M. Jamil, D. Momeni and R. Myrzakulov, {\it Chin. Phys. Lett.} {\bf29}, 109801 (2012).
\item[]65. F.G. Alvarenga, M.J. Houndjo, A.V. Monwanou and J.B. Chabi Oron, {\it J. Mod.  Phys.} {\bf 4}, 130 (2013)  [arXiv,gr-qc/1205.4678].
\item[]66. M. Sharif, S. Rani and R. Myrzakulov, {\it Eur. Phys. J. Plus} {\bf 128}, 123 (2013)  [arXiv,gr-qc/1210.2714].
\item[]67. M. Sharif and M. Zubair, {\it J. Phys. Soc. Jpn.} {\bf 82}, 014002 (2013)  [arXiv,gr-qc/1210.3878].
\item[]68. S. Chakraborty, {\it Gen. Rel. Grav.} {\bf45}, 2039 (2013)  [arXiv,gen-ph/1212.3050].
\item[]69. F.G. Alvarenga, A. de la Cruz-Dombriz, M.J.S. Houndjo, M. E. Rodrigues and D. S$\acute{a}$ez-G$\acute{o}$mez, {\it Phys. Rev. D } {\bf 87}, 103526 (2013)   [arXiv,gr-qc/1302.1866].
\item[]70. A. Pasqua, S. Chattopadhyay and I. Khomenkoc, {\it Canadian J. Phys.} {\bf 91}, 632 (2013).
\item[]71. H. Shabani and M. Farhoudi, {\it Phys. Rev. D} {\bf 88}, 044048 (2013).
\item[]72. M. Sharif and M. Zubair, {\it J. Phys. Soc. Jpn.} {\bf 82}, 064001 (2013)  [arXiv,gr-qc/1310.1067].
\item[]73. C.P. Singh and V. Singh, {\it Gen. Rel. Grav.} {\bf 46}, 1696 (2014).
\item[]74. K.S. Adhav, {\it Astrphys. Space. Sci.} {\bf 339}, 365 (2013).
\item[]75. S.D. Katore and A.Y. Shaikh, {\it Prespacetime J.} {\bf3}, 1087 (2012).
\item[]76. M. Sharif and M. Zubair, {\it J. Phys. Soc. Jpn.} {\bf 81}, 114005 (2012)  [arXiv,gr-qc/1301.2251].
\item[]77. V.U.M. Rao and D. Neelima, {\it Astrphys. Space. Sci.} {\bf 345},  427 (2013).
\item[]78. N. Ahmed and A. Pradhan, {\it Int. J. Theor. Phys.} {\bf53}, 289 (2014).
\item[]79. I. Prigogine, J. Gohenian, E. Gunzig and P. Nardone, {\it Gen. Rel. Grav.} {\bf 21}, 767 (1989).

\end{description}}

\end{document}